\documentclass[aps,reprint,nofootinbib,nobibnotes,showpacs,superscriptaddress]{revtex4-1}

\usepackage[english]{babel}
	\addto\captionsenglish{%
	}

\usepackage{graphicx}

\usepackage{amsmath}
\usepackage{amssymb}
\usepackage[version=3]{mhchem}	
\usepackage{bm}	
\usepackage{nicefrac}	
\usepackage{mathrsfs}

\usepackage[hidelinks]{hyperref}
\hypersetup{
    colorlinks,
    linkcolor={red!70!black},
    citecolor={green!70!black},
    urlcolor={blue!70!black}
}

\usepackage[usenames,dvipsnames]{xcolor}	

\newcommand{\bb}{$0\nu \beta \beta$} 
\newcommand{\vvbb}{$2\nu \beta \beta$} 

\newcommand{\mbb}{m_{\beta \beta}} 
\newcommand{\mbbM}{\mbb^{\max}} 
\newcommand{\mbbm}{\mbb^{\min}} 

\newcommand{\taubb}{\mbox{\normalsize $t^{\mbox{\tiny $\nicefrac{1}{2}$}}$}} 
\newcommand{\taubbiso}[1]{\mbox{\normalsize $t^{\mbox{\tiny $\nicefrac{1}{2}$}}_{\mbox{\scriptsize #1}}$}} 
\newcommand{\Gi}{\mathcal{G}_{0\nu,i}} 

\newcommand{\NH}{$\mathcal{NH}$} 
\newcommand{\IH}{$\mathcal{IH}$} 

\newcommand{\meV}{\text{meV}}	\newcommand{\eV}{\text{eV}}

\newcommand{\yr}{\text{yr}}	

\newcommand{\el}{\text{e}}	

\newcommand{\theor}{\mbox{\scriptsize(th.)}}		
\newcommand{\exper}{\mbox{\scriptsize(exp.)}}		

\newcommand{\xcCL}{\mbox{\tiny(90\% C.\,L.)}}		
\newcommand{\zb}{\mbox{\tiny(zero bkg)}}		

\newcommand{\meas}{\mbox{\tiny meas}}		

\begin{document}

\title{New expectations and uncertainties on neutrinoless double beta decay}

	\author{Stefano Dell'Oro}
		\email{\ttfamily stefano.delloro@gssi.infn.it}
	\author{Simone Marcocci}
		\email{\ttfamily simone.marcocci@gssi.infn.it}
		\affiliation{Gran Sasso Science Institute, Viale Crispi 7, 67100 L’Aquila, Italy \\}%
	\author{Francesco Vissani}
		\email{\ttfamily francesco.vissani@lngs.infn.it}
		\affiliation{INFN, Laboratori Nazionali del Gran Sasso, Assergi, L’Aquila, Italy \\}%
		\affiliation{Gran Sasso Science Institute, Viale Crispi 7, 67100 L’Aquila, Italy \\}%

\date{\today}
	

	\begin{abstract}
		The hypothesis that the Majorana mass of ordinary neutrinos dominates the rate of neutrinoless double beta decay 
		is investigated. Predictions from neutrino oscillations are updated. Nuclear uncertainties are discussed, 
		evaluating the impact of the quenching of the axial vector coupling constant in the nuclear medium, recently 
		pointed out by Iachello \emph{et al.}~[Phys.\ Rev.\ C\,87, 014315 (2013)]. 
		Also, the sensitivity of present and future experiments is assessed 
		and possible implications of the knowledge on neutrino masses from cosmology are studied. The predictions from 
		neutrino oscillations are compared with the results from cosmology and from neutrinoless double beta decay 
		searches, emphasizing the important role of the measurement errors. The obstacles to an experimental 
		determination of the Majorana phases are pointed out. 
	\end{abstract}

	\pacs{14.60.Pq,
		23.40.-s
		\hspace{145pt}%
		DOI: \href{http://journals.aps.org/prd/abstract/10.1103/PhysRevD.90.033005}{10.1103/PhysRevD.90.033005}}


	\maketitle


\section{Introduction}

	The search for neutrinoless double beta decay (\bb) probes lepton number conservation 
	and allows us to investigate the nature of the neutrino mass eigenstates.
	In this work, we perform an updated and in-depth study of the conservative assumption that this transition is due 
	to the exchange of the three known neutrinos, endowed with Majorana mass. We emphasize  
	the role of the uncertainties due to the quenching of the axial vector coupling constant in the nuclear medium,
	recently pointed out by Iachello \emph{et al.}~\cite{Barea&al:2013}.
	
	First, we update the predictions from oscillations (Sec.\ \ref{sec:mbb}). We compare these predictions with 
	the recent experimental results on \bb~in Sec.\ \ref{compon}. The sensitivity of future experiments, 
	defined by taking into account the uncertainties, is discussed in Sec.\ \ref{sec:fut_exp}.
	In Sec.\ \ref{sec:cosm_bounds}, we analyze the implication of recent bounds and hints for the neutrino mass 
	obtained in cosmology. Finally, in Sec.\ \ref{uevos}, we discuss whether it could be possible to measure the 
	\emph{Majorana phases} and/or discriminate the two neutrino mass hierarchies while quantifying
	the role of the errors of measurement.


\section{Updated predictions from oscillations}
	\label{sec:mbb}

	\begin{figure*}[htb]
		\centering
		\includegraphics[width=\columnwidth]{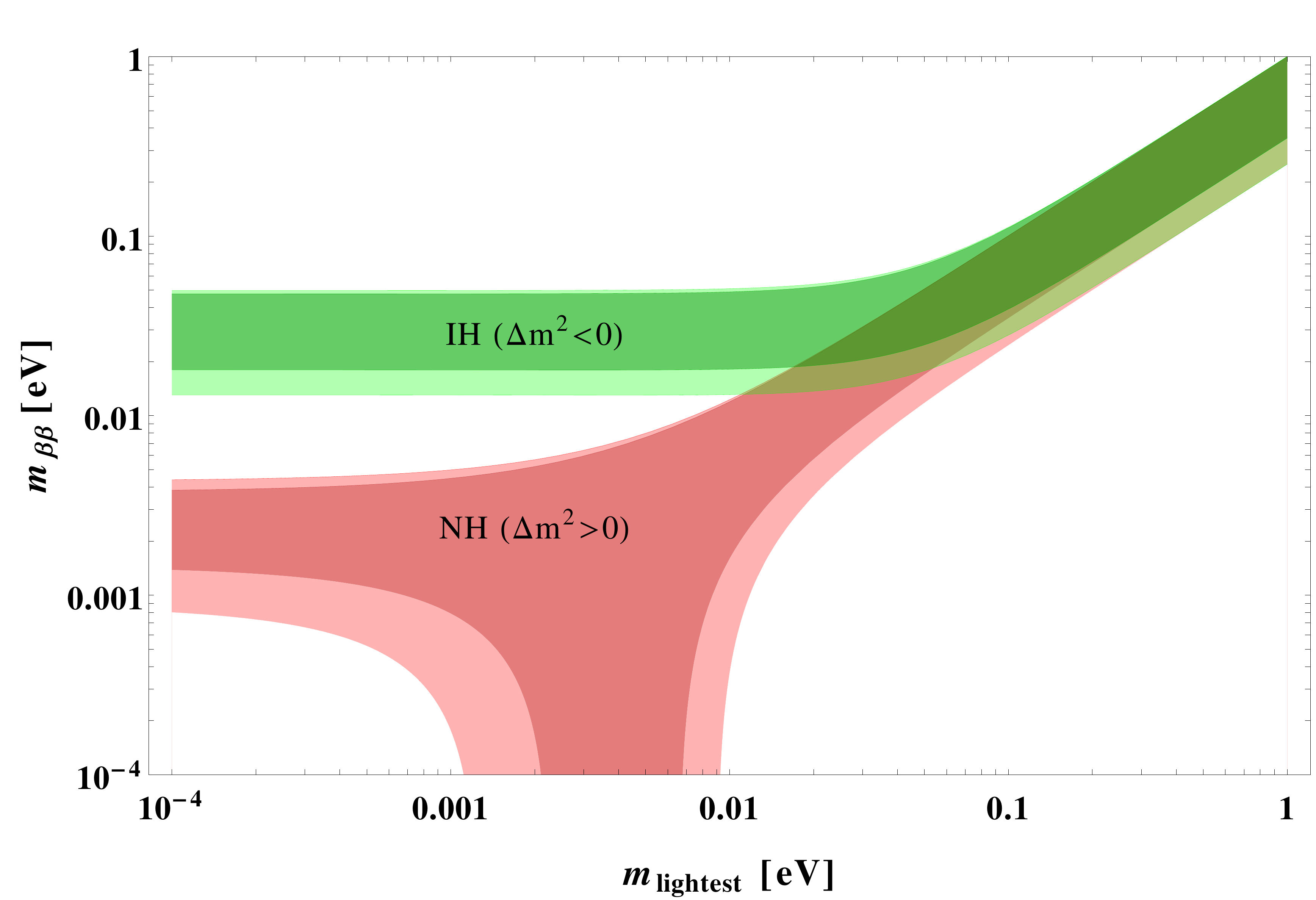} \label{fig:DBD_graph}
		\includegraphics[width=\columnwidth]{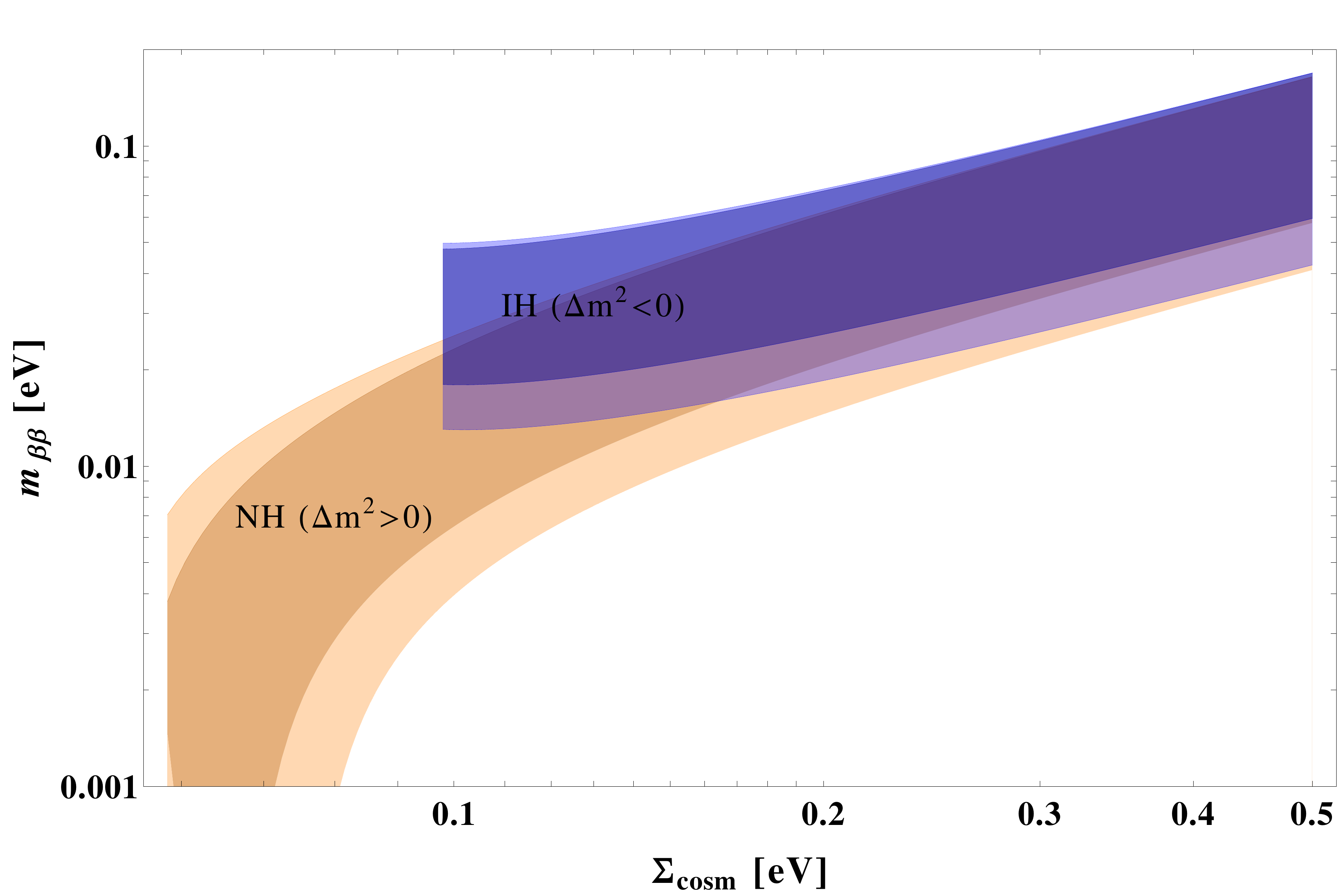} \label{fig:DBD_graph_cosm}
		\caption{Updated predictions on $\mbb$ from oscillations as a function of the lightest neutrino mass (left) 
			and of the cosmological mass (right) in the two cases of \NH~and \IH.
			The shaded areas correspond to the $3\sigma$ regions due to error propagation of the uncertainties on the 
			oscillation parameters.}
		\label{fig:DBD_graph}
	\end{figure*}

	Assuming that the \bb~transition is caused by the exchange of \emph{ordinary} neutrinos, the key parameter that 
	regulates its rate is the \emph{Majorana effective mass}, namely,
	\begin{equation} \label{eq:mbbDef}
		\mbb \equiv \biggl| \el^{i\alpha_1}|U_{ei}^2|m_1 + \el^{i\alpha_2}|U_{e2}^2|m_2 + |U_{e3}^2|m_3 \biggr|.
	\end{equation}
	It represents the absolute value of the ee entry of the neutrino mass matrix. Here, $m_i$ are the masses 
	of the individual neutrinos $\nu_i$, $\alpha_{1,2}$ are the Majorana phases and 
	$U_{ei}$ are the elements of the mixing matrix that define the composition of
	the electron neutrino: $|\nu_\el \rangle = \sum_{i=1}^3 U_{ei}^* |\nu_i \rangle$.

	The present information on three-flavor neutrino oscillations is compatible with two different neutrino mass spectra: 
	\emph{normal hierarchy} (\NH) and \emph{inverted hierarchy} (\IH). 
	In the former case the mass-squared difference between the two heavier states is much
	larger than the one between the two lighter states. In the latter case, the opposite is true.

	Thanks to the knowledge of the oscillation parameters, it is possible to constrain the parameter $\mbb$.
	However, since the complex phases $\alpha_{1,2}$ in Eq.\ (\ref{eq:mbbDef}) cannot be probed by oscillations and 
	are unknown, the allowed region
	for $\mbb$ is obtained letting them vary freely. The expressions for the resulting extremes are, \cite{Vissani:1999}:
	\begin{align}\label{talp1}
			&\mbbM = \sum_{i=1}^3 \bigl| U_{ei}^2 \bigr| m_i \\
			&\mbbm = \max \Bigl\{ 2 \bigl| U_{ei}^2 \bigr| m_i - \mbbM, 0 \Bigr\} \quad i=1,2,3 . \label{talp2}
	\end{align}
	We adopt the graphical representation of $\mbb$ introduced in~\cite{Vissani:1999} and refined 
	in~\cite{Feruglio&al:2002,Vissani&Strumia:2006}. 
	It consists in plotting $\mbb$~in bilogarithmic scale as a function of the mass of the lightest neutrino, 
	both for the cases of \NH~and of \IH. The resulting plot, according to the new values of the oscillation 
	parameters in~\cite{Capozzi&al:2013}, is shown in the left panel of Fig.\ \ref{fig:DBD_graph}.
	The uncertainties on the various parameters entering 
	Eqs.\ (\ref{talp1}) and (\ref{talp2}) are propagated using the procedures described in the Appendix, 
	(Eq.\ (\ref{eq:err_prop})).
	This results in a wider allowed region, which corresponds to the shaded parts in the picture.
	
	It is also useful to express the parameter $\mbb$ as a function of a directly observable parameter, rather than as 
	a function of the lightest neutrino mass. 
	A natural choice is the cosmological mass $\Sigma$, defined as the sum of the three active neutrino masses 
	($\Sigma \equiv m_1+m_2+m_3$). The close connection between the neutrino masses measurements obtained in the 
	laboratory and those probed by cosmological observations was outlined long ago~\cite{Zeldovich&Khlopov:1981}.
	Furthermore, the measurements of $\Sigma$ have recently reached important sensitivities, as discussed below. 
	For these reasons, 
	we also update the plot of the dependence of the Majorana effective mass $\mbb$ on the cosmological mass $\Sigma$, 
	using the representation originally introduced in~\cite{Fogli&al:2004}.

	From the definition of $\Sigma$, we can write:
	\begin{equation}
		\Sigma = m_l + \sqrt{m_l^2 + a^2} + \sqrt{m_l^2 + b^2}
		\label{eq:imhotep}
	\end{equation}
	where $m_l$ is the mass of the lightest neutrino and $a$ and $b$ are different constants depending on the neutrino mass 
	hierarchy. Through Eq.\ (\ref{eq:imhotep}) one can establish a direct relation between $\Sigma$ and $m_l$ and thus
	it is straightforward to plot $\mbb$ as a function of $\Sigma$. 
	Concerning the treatment of the uncertainties, we use again the assumption 
	of Gaussian fluctuations and the prescription reported in the Appendix.
	The result of the plotting in this case is shown in the right panel of Fig.\ \ref{fig:DBD_graph}. 
	
\section{Comparison with the experimental results}
	\label{compon}

	\vspace{-30pt}

	\begin{center}
	\begin{table*}[t]
	\caption{Lower bounds achievable for $\mbb$ by some \bb~experiments, depending on their reached sensitivities
		(upper group) or sensitivity goals (lower group). 
		The different results correspond to the different quenching of $g_A$, according to the definitions in 
		Eq.\ (\ref{cazzes}). The $1\sigma$
		uncertainties on $\mbb$ are calculated by assuming uncertainties both on the matrix elements and phase space factors, 
		according to~\cite{Barea&al:2013} and~\cite{Kotila&Iachello:2012} respectively.}
	\small{
	\begin{ruledtabular}
	\begin{tabular}{l c c c c c c}
		Experiment	&Isotope	&$\taubb   \xcCL \, [10^{25}\,\yr] \quad$ 	
		&\multicolumn{3}{c}{Lower bound for $\mbb\, [\eV]$} \\
		\cline{4-6}
					&			&			&$g_{\mbox{\tiny nucleon}}$	&$g_{\mbox{\tiny quark}}$	&$g_{\mbox{\tiny phen.}}$		\\
		\hline
		IGEX,\,\cite{sens_IGEX}						&\ce{^{76}Ge}		&$1.57$	&$0.31\pm0.03$	&$0.49\pm0.05$	&$1.44\pm0.16$	\\
		HEIDELBERG-MOSCOW,\,\cite{sens_HDM}		&\ce{^{76}Ge}		&$1.9	$	&$0.28\pm0.03$	&$0.44\pm0.05$	&$1.31\pm0.14$	\\
		GERDA-I,\,\cite{sens_GERDA-I}				&\ce{^{76}Ge}		&$2.1 $	&$0.26\pm0.03$	&$0.42\pm0.05$	&$1.25\pm0.14$	\\
		KamLAND-Zen-I,\,\cite{sens_KamLAND-Zen}	&\ce{^{136}Xe}		&$1.9 $	&$0.18\pm0.02$	&$0.29\pm0.03$	&$1.06\pm0.12$	\\
		KamLAND-Zen-II,\,\cite{KamLAND-Zen_Neutrino}	&\ce{^{136}Xe}		&$1.3 $	&$0.22\pm0.02$	&$0.35\pm0.04$	&$1.28\pm0.14$	\\
		EXO-200,\,\cite{sens_EXO_new}				&\ce{^{136}Xe}		&$1.1 $	&$0.24\pm0.03$	&$0.38\pm0.04$	&$1.39\pm0.15$	\\
		Combined \ce{Ge},\,\cite{sens_GERDA-I}	&\ce{^{76}Ge}		&$3.0 $	&$0.22\pm0.02$	&$0.35\pm0.04$	&$1.05\pm0.11$	\\
		Combined \ce{Xe}								&\ce{^{136}Xe}		&$2.6 $	&$0.15\pm0.02$	&$0.25\pm0.03$	&$0.91\pm0.10$	\\
		Combined \ce{Ge}+\ce{Xe}				&\ce{^{76}Ge}/\ce{^{136}Xe}&	&$0.15\pm0.01$	&$0.24\pm0.02$	&$0.81\pm0.07$	\\
		\\[-2.0mm]
		CUORE,\,\cite{sens_CUORE}					&\ce{^{130}Te}		&$9.5 $	&$0.07\pm0.01$	&$0.11\pm0.01$	&$0.39\pm0.04$	\\
		GERDA-II,\,\cite{sens_GERDA}				&\ce{^{76}Ge}		&$15 $	&$0.10\pm0.01$	&$0.16\pm0.02$	&$0.47\pm0.05$	\\
		SuperNEMO,\,\cite{sens_SuperNEMO}		&\ce{^{82}Se}		&$10 $ 	&$0.07\pm0.01$	&$0.12\pm0.01$	&$0.36\pm0.04$	\\
		\end{tabular}
		\end{ruledtabular}
		}
	\label{tab:mbb_bounds}
	\end{table*}
	\end{center}

\subsection{Experimental bounds}
	\label{sec:exp}

	Recently, several experiments have obtained bounds on $\taubb\,\exper$ above $10^{25}$\yr. 
	The results are summarized in the upper part of Table \ref{tab:mbb_bounds}. 
	They were achieved thanks to the study of two nuclei: \ce{^{76}Ge} and \ce{^{136}Xe}. 
	The 90\% C.\,L.\ bound from \ce{^{76}Ge}, obtained by combining GERDA-I, Heidelberg-Moscow and 
	IGEX via the recipe of Eq.\ (\ref{eq:comb_t}), $3.2\cdot 10^{25}\,\yr$, is almost identical to the one 
	quoted by the GERDA Collaboration, $3.0\cdot 10^{25}\,\yr$, \cite{sens_GERDA-I}.
	By combining the first KamLAND-Zen results on \bb~(namely, KamLAND-Zen-I~\cite{sens_KamLAND-Zen}), 
	and the new ones obtained after the scintillator 
	purification (KamLAND-Zen-II~\cite{KamLAND-Zen_Neutrino}), the same procedure gives $2.3\cdot 10^{25}\,\yr$, 
	which differs a little bit from the combined limit 
	quoted by the Collaboration~\cite{KamLAND-Zen_Neutrino}, $2.6\cdot 10^{25}\,\yr$. 
	When we combine the two results of KamLAND-Zen and the one from EXO-200 using again 
	the procedure of Eq.\ (\ref{eq:comb_t}), we get $2.6\cdot 10^{25}$ yr, which is equal to the KamLAND-Zen limit alone. 
	In view of the above discussion and in order to be as conservative as possible, we will adopt as combined 
	90\% C.\,L.\ bounds the following values:
	\begin{equation}
	 	\taubbiso{\ce{Ge}} > 3.0 \cdot 10^{25}\,\yr 
	 	\quad \mbox{and} \quad
	 	\taubbiso{\ce{Xe}} > 2.6 \cdot 10^{25}\,\yr.
	\label{eq:summa}
	\end{equation}
	More experiments are also expected to produce important new results in the coming years. A few selected ones are also 
	reported in the lower part of Table \ref{tab:mbb_bounds}.

\subsection{Nuclear physics and \bb}
	\label{sec:nucl_asp}

	Assuming that the transition is dominated by the exchange of ordinary neutrinos with Majorana mass, the theoretical 
	expression of the half-life in an $i$th experiment based on a certain nucleus is:
	\begin{equation}
		\taubbiso{$i$}\theor = \frac{m_\el^2}{\Gi \, \mathscr{M}_i^2\, \mbb^2}
	\label{eq:tau}
	\end{equation}
	where $m_\el$ is the electron mass, 
	$\Gi$ the phase space factor (usually given in inverse years) and $\mathscr{M}_i$ the nuclear matrix element, 
	an adimensional quantity of enormous importance. In recent works, this last term 
	is written emphasizing the axial coupling $g_A$:
	\begin{equation}
		\mathscr{M}_i = g_A^2 \cdot \mathscr{M}_{0\nu,i}.
	\end{equation}
	$\mathscr{M}_{0\nu,i}$ depends mildly on $g_A$ and can be evaluated by theoretically modeling the nucleus. 
	This is independent on $g_A$ if the same quenching is assumed both for the 
	vector and axial coupling constants, as we do here for definiteness, 
	following~\cite{Barea&al:2013} (as discussed in the reference, some residual dependence upon $g_A$ could 
	be attributed to a different 	renormalization of the two coupling constants).
	On the contrary, $\Gi$ is a constant parameter, independent on $g_A$, and it is reasonably well known. Its value 
	can be found, e.\,g., in~\cite{Kotila&Iachello:2012} for all the candidate \bb~emitters. 
	
	As a consequence, an experimental limit on the half-life translates into a limit on 
	the mass parameter:
	\begin{equation}
		\mbb \le \frac{m_\el}{\mathscr{M}_i\  \sqrt{\Gi \, \taubbiso{$i$}\exper}}.
	\label{eq:mbb_bound}
	\end{equation}
	The main sources of uncertainties in the inference are the nuclear matrix elements.
	The first calculations of $ \mathscr{M}_{0\nu,i}$ that also estimated the errors, based on the ``QRPA'' 
	description of the nucleus, assessed 
	a relatively small intrinsic error of $\sim 20\%$~\cite{Rodin&al:2003,Rodin&al:2006}. 
	The validity of these conclusions has been recently supported 
	by a completely independent calculation based on the ``IBM2'' description of the 
	nucleus~\cite{Kotila&Iachello:2012,Barea&al:2013}. 
	
	However, the same papers have also emphasized a more important role 
	of the axial coupling $g_A$ than originally thought. In other words, the real theoretical issue concerns $\mathscr{M}_{i}$.  
	Indeed, it is commonly expected that the value $g_A\simeq 1.269$ 
	measured in the weak interactions and decays of nucleons is modified (or, \emph{renormalized}) 
	in the nuclear medium toward the value appropriate for quarks~\cite{Rodin&al:2003,Rodin&al:2006,Simkovic&al:2013b};  
	the plausibility of further modification (reduction) has been argued in~\cite{Barea&al:2013}, based on the 
	knowledge on the double beta decay with neutrinos (\vvbb).
	In light of this discussion, a conservative treatment of the uncertainties should consider at least three cases: 
	\begin{equation}
	g_A~=~
		\begin{cases} \label{cazzes}
			g_{\mbox{\tiny nucleon}} &= 1.269 \\
			g_{\mbox{\tiny quark}}   &= 1 \\
			g_{\mbox{\tiny phen.}}   &= 1.269 \cdot A^{-0.18}.
		\end{cases}
	\end{equation}
	We will refer to the last formula with the name \emph{maximal quenching}. It includes phenomenologically the 
	effect of the atomic number $A$. The $g_{\mbox{\tiny phen.}}$ parametrization as a function of $A$ comes directly from 
	the comparison between the theoretical half-life for \vvbb~and its observation in different nuclei~\cite{Barea&al:2013}. 
	
	Needless to say, the validity of the assumption that the quenching is the same both 
	for the \vvbb~and the \bb~cases is still an open issue. We stress that this is just a phenomenological description of 
	the quenching, since the specific behavior is different in each nucleus and it somewhat differs from 
	this parametrization~\cite{Barea&al:2013}. 
	Nonetheless, the assumption described in Eq.\ (\ref{cazzes}) seems a reasonable one and deserves discussion in the 
	present context. 
	In fact, the question of \emph{which is the true value of $g_A$} introduces a considerable uncertainty in the 
	inferences concerning massive neutrinos. 

	\begin{center}
	\begin{table*}[t]
	\caption{Sensitivity and exposure necessary to discriminate between \NH~and \IH: the goal is $\mbb=8\,\meV$. The two cases 
		refer to the unquenched value of $g_A=g_{\mbox{\tiny nucleon}}$ (mega) and $g_A=g_{\mbox{\tiny phen.}}$ 
		(ultimate). The calculations are performed assuming \emph{zero background} experiments with $100\%$ detection 
		efficiency and no fiducial volume cuts. The last column shows the maximum value of the product $B \cdot \Delta$ in 
		order to actually comply with the zero background condition.}
	\small{
	\begin{ruledtabular}
	\begin{tabular}{l c c c c}
		Experiment		&Isotope			&$\taubb \, [\yr] \quad$		&\multicolumn{2}{c}{Exposure (estimate)} \\
		\cline{4-5}
		&	&	&$M \cdot T$ \,[ton$\cdot$yr]		 &$B \cdot \Delta \, \zb$\,[counts/kg/yr]					\\
		\hline
		Mega Te			&\ce{^{130}Te}		&$6.8 \cdot 10^{27}$	&$2.1$	&$4.7\cdot10^{-4}$				\\
		Mega Ge			&\ce{^{76}Ge}		&$2.3 \cdot 10^{28}$	&$4.1$	&$2.4\cdot10^{-4}$				\\
		Mega Xe			&\ce{^{136}Xe}		&$9.7 \cdot 10^{27}$	&$3.2$ 	&$3.2\cdot10^{-4}$				\\
		\\[-2.0mm]
		Ultimate Te		&\ce{^{130}Te}		&$2.3 \cdot 10^{29}$	&$71$		&$1.4\cdot10^{-5}$				\\
		Ultimate Ge		&\ce{^{76}Ge}		&$5.1 \cdot 10^{29}$	&$93$		&$1.1\cdot10^{-5}$				\\
		Ultimate Xe		&\ce{^{136}Xe}		&$3.3 \cdot 10^{29}$	&$109$	&$9.2\cdot10^{-6}$				\\
		\end{tabular}
		\end{ruledtabular}
		}
	\label{tab:exposure}
	\end{table*}
	\end{center}

	\vspace{-30pt}

\subsection{Sensitivity of present experiments}

	Once the experimental limits on the half-lives are known, by using the phase space of~\cite{Kotila&Iachello:2012} 
	and the matrix elements of~\cite{Barea&al:2013},
	it is possible to find the lower bounds on $\mbb$ according to Eq.\ (\ref{eq:mbb_bound}).
	Table \ref{tab:mbb_bounds} shows the results for the experiments considered in 	Sec.\ \ref{sec:mbb}.
	In order to obtain the combined bounds, the procedure shown in the Appendix was used.
	The different values of $\mbb$ correspond to the three quenching scenarios considered in 
	Sec.\ \ref{sec:nucl_asp}. The $1\sigma$ errors
	of Table \ref{tab:mbb_bounds} were computed according to Eq.\ (\ref{eq:err_prop}), assuming both uncertainties on the 
	matrix elements and phase space factors, as reported in~\cite{Barea&al:2013} and~\cite{Kotila&Iachello:2012} 
	respectively. Nonetheless, the former error gives the main contribution.
	
	The importance of the $g_A$ quenching is evident from 
	the table: the sensitivity for the same experiment 
	in the two cases of $g_{\mbox{\tiny nucleon}}$ and $g_{\mbox{\tiny phen.}}$ differs by a factor $\sim 5$.
	This is graphically shown in Fig.\ \ref{fig:mbb_bande}, where the present best limit on $\mbb$ coming from the 
	combination \ce{Ge}+\ce{Xe} (obtained as described in  the Appendix) is plotted in correspondence of the 
	considered values of $g_A$.

\section{Sensitivity of Future Experiments}
	\label{sec:fut_exp}

	Now we consider a next generation experiment (call it a \emph{mega} experiment) and a next-to-next 
	generation one (an \emph{ultimate} experiment) with enhanced sensitivity. First of all, we should clarify which is the 
	physics goal that we would like to achieve.
	
	Plausibly, the most honest way to talk of the sensitivity is in terms of exposure or of half-life time that can be 
	probed. From the point of view of the physical interest, however, besides the hope of 
	discovering the \bb, the most exciting investigation 
	that can be imagined at present is the exclusion of the \IH~case. 
	This is the goal that most of the experimentalists are trying to reach with \bb~experiments, working 
	in the above assumptions and supposing  that \bb~will not be found. 
	For this reason, we require a sensitivity:
	\begin{equation} \notag
		\mbb = 8\,\meV.
	\end{equation}
	The \emph{mega} experiment is the one that  satisfies this requirement in the most favorable case, namely, 
	when the quenching of $g_A$ is absent. Instead, the \emph{ultimate} experiment assumes that $g_A$ is maximally 
	quenched. We chose the 8 meV value because, even taking into account the residual uncertainties on the nuclear 
	matrix elements, the overlap with the allowed band for $\mbb$ in the \IH~is excluded. 
	In fact, the uncertainties on Ge and Xe nuclei amount to $\sim 20\%$, as discussed above.
	Notice that we are assuming that at some point the issue of the quenching will be sorted out.
	Through Eq.\ (\ref{eq:mbb_bound}), 
	we obtain the corresponding value of $\taubb$ and thus we calculate the needed exposure to accomplish the task.

	The law of radioactive decay prescribes that
	\begin{equation}
			\taubb = \ln 2 \cdot T \cdot \varepsilon \cdot \frac{x\cdot \eta\cdot N_{\mbox{\tiny A}} \cdot M}%
			{\mathcal{M}_A \cdot N_{\mbox{\tiny S}}}
		\label{eq:mbb_seen}
	\end{equation}
	where $T$ is the measuring time, $\varepsilon$ is the detection efficiency, $x$ is the stoichiometric multiplicity of 
	the element containing the $\beta \beta$ candidate, $\eta$ is the $\beta \beta$ candidate isotopic abundance, 
	$N_{\mbox{\tiny A}}$ is the Avogadro number, $\mathcal{M}_A$ is the compound molar mass and 
	$N_{\mbox{\tiny S}}$ is the number of 
	observed decays in the region of interest.
	Let us focus on the optimal experimental condition, when the contribution of the background counts is negligible
	(\emph{zero background} condition). This means that we require:
	\begin{equation}
		M\cdot T\cdot B\cdot \Delta \lesssim 1
	\label{eq:zb}
	\end{equation}
   where $M$ is the detector mass; $B$ is the background level per unit mass, energy and time; and $\Delta$ is the 
	full width half maximum (FWHM) energy resolution. 
	Now, if we suppose $\varepsilon \simeq 1$ (detector efficiency of $100\%$ and no fiducial volume cuts), 
	$x\simeq\eta\simeq1$ (all the mass is given by the candidate nuclei), and we assume one observed event 
	(i.\,e.\ $N_{\mbox{\tiny S}} = 1$) in the region of interest, Eq.\ (\ref{eq:mbb_seen}) simplifies to
	\begin{equation} 
		M \cdot T =\frac{\mathcal{M}_A \cdot \taubb}{\ln 2 \cdot N_{\mbox{\tiny A}}} . 
	\label{eq:our_sensitivity}
	\end{equation}
	This is the equation we used to estimate the product $M\cdot T$ (exposure), 
	and thus to assess the sensitivity of the mega and ultimate scenarios.
	The key input is, of course, the theoretical expression of $\taubb$.
	The calculated values of the exposure 
	are shown in Table \ref{tab:exposure} for the 
	three considered nuclei: \ce{^{76}Ge}, \ce{^{130}Te} and \ce{^{136}Xe}.
	The last column of the table gives 
	the maximum allowed value of the product $B\cdot \Delta$ that satisfies  Eq.~(\ref{eq:zb}).

	Finally, Fig.\ \ref{fig:mbb_bande} shows the present knowledge on \bb~according to the best combined limit of Ge+Xe 
	of Table \ref{tab:mbb_bounds} compared to the mega/ultimate scenarios. 
	We report the three possible predictions on the bounds on $\mbb$ according to the three quenching scenarios considered.
	The presence of $1\sigma$ bands instead of single lines is due to the propagation of the residual uncertainties 
	on the nuclear matrix elements.
	The mega/ultimate scenarios are presented as the gray band.
	
\section{Implications of cosmology}
	\label{sec:cosm_bounds}

	\begin{figure}[t]
		\centering
		\includegraphics[width=\columnwidth]{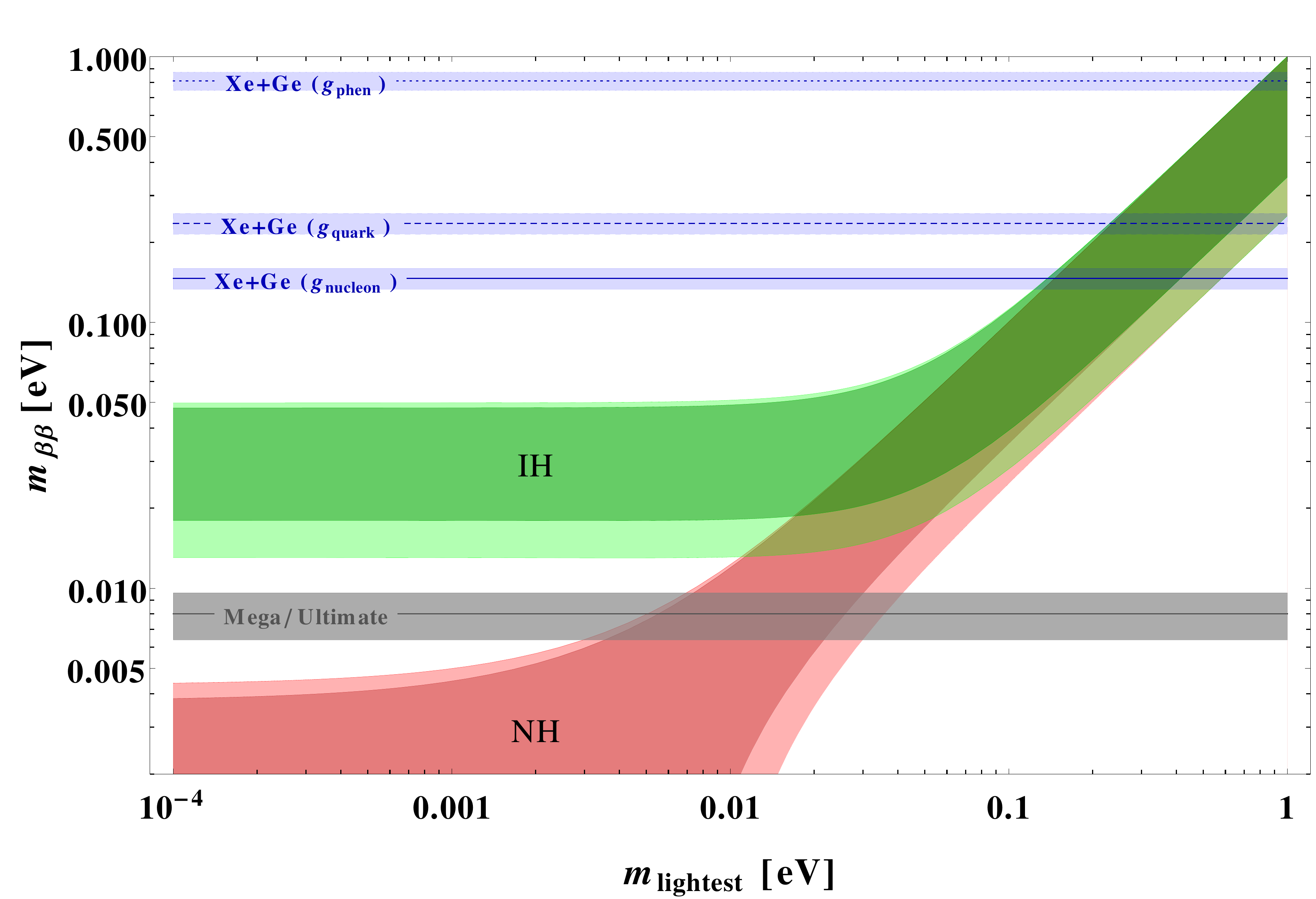}
		\caption{Present sensitivity on $\mbb$, according to the Ge+Xe combined limit, in the three quenching scenarios. 
			The blue ($1\sigma$) bands come from the uncertainties on the nuclear matrix elements. The gray band at 
			$\mbb=(8 \pm 1.6)\,\meV$ concerns the ultimate and mega experiments, discussed in the text.}
		\label{fig:mbb_bande}
	\end{figure}
	
	Here we discuss the possibility of taking advantage of the knowledge about the neutrino cosmological mass to make 
	inferences on some \bb~experiment results (or expected ones). 
	We consider only the optimistic assumption that $g_A$ is unquenched. The changes induced 
	by the quenching can be easily understood by considering, e.\,g., its impact in Fig.\ \ref{fig:mbb_bande}. 
	Evidently, this weakens the reach of each experiment, rescaling the possible bounds or measurements toward 
	larger values.
	
\subsection{Information from cosmology}

	The new experimental limit provided by the Planck 
	experiment on $\Sigma$ is 0.23\,eV at 95\% C.\,L.~\cite{Ade&al:2013}. 
	Interestingly, several studies have emphasized some tension 
	between the data of Planck and those from galaxy counts and lensing. 
	Their combination suggests a nonzero best fit value of the mass, in the range (0.3--0.4)\,eV and with an error 
	of about 30\%~\cite{Wyman&al:2014,Battye&Moss:2014}.
	Taking into account these data, we will consider two scenarios for the subsequent discussion:
	\begin{equation}
		\begin{split}
			&\Sigma < 0.19\,\eV \,(90\% \, \text{C.\,L.}) \quad \mbox{ (conservative), \cite{Ade&al:2013}} \\[+5pt]
			&\Sigma = (0.320 \pm 0.081)\,\eV \quad \, \mbox{ (aggressive), \cite{Battye&Moss:2014}}.
		\end{split}
		\label{eq:limits_sigma}
	\end{equation}
	We consider these two cases since, at present, the evidence for nonzero neutrino masses is not strong and the 
	possibility of unexpected systematics cannot be excluded.
	
	The results on $\Sigma$ can provide us precious information on \bb~in the assumption that this transition is 
	dominated by the light neutrinos exchange.
	For example, by looking at Fig.\ \ref{fig:DBD_graph} and in the conservative limit in 
	Eq.\ (\ref{eq:limits_sigma}), it seems useless to look for \bb~with an experiment with a 
	sensitivity on $\mbb$ of $100\,\meV$ or more.
	If, instead, the claim of measurement in the equation is correct, and as soon as \bb~experiments  
	probe the transition rate, we will obtain information on the quenching factor. A successful measurement in the next 
	generation of experiments, for example, would mean that the quenching is reduced or absent.

	\begin{figure*}[t]
		\centering
		\includegraphics[width=\columnwidth]{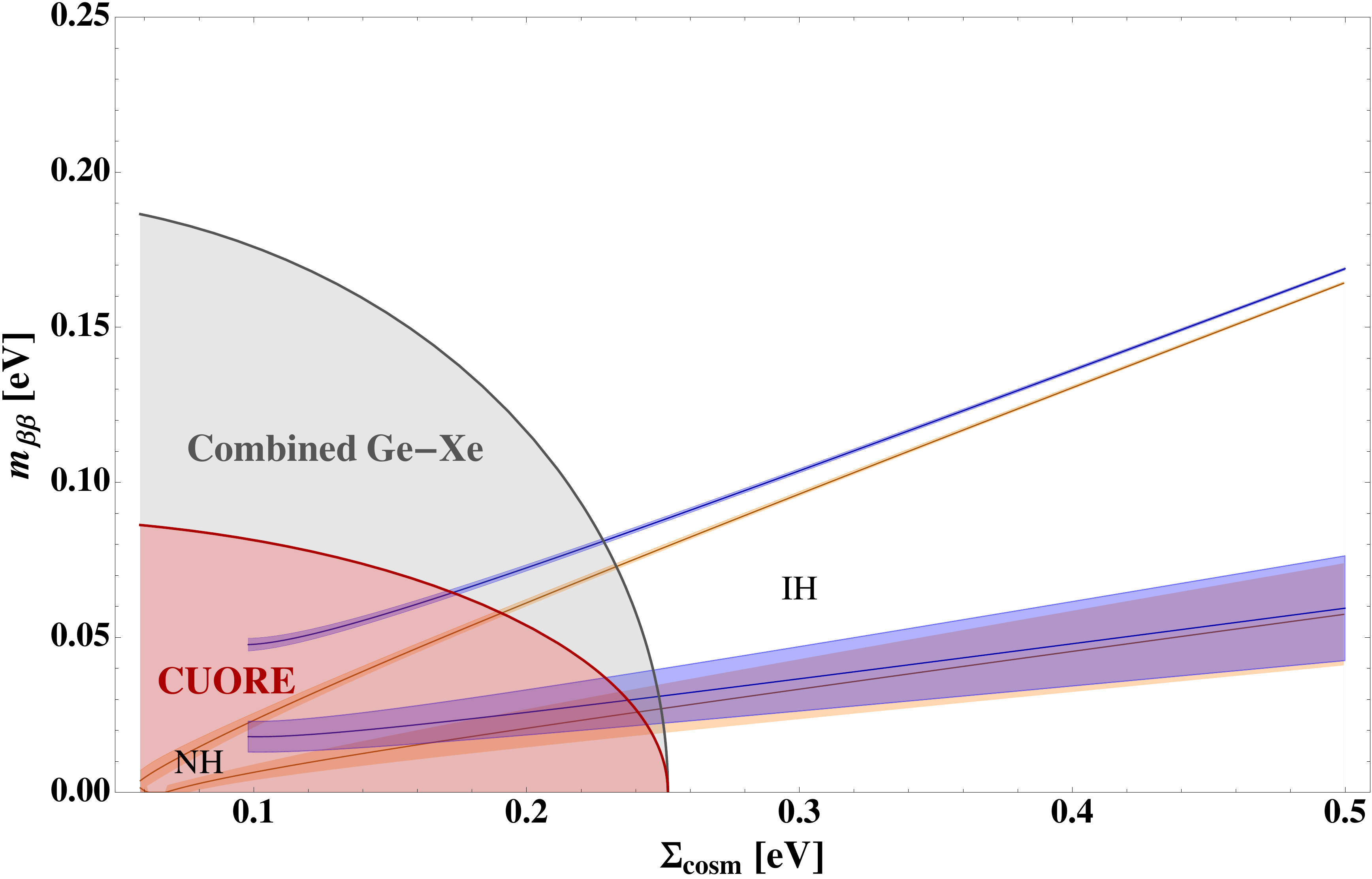} 
		\includegraphics[width=\columnwidth]{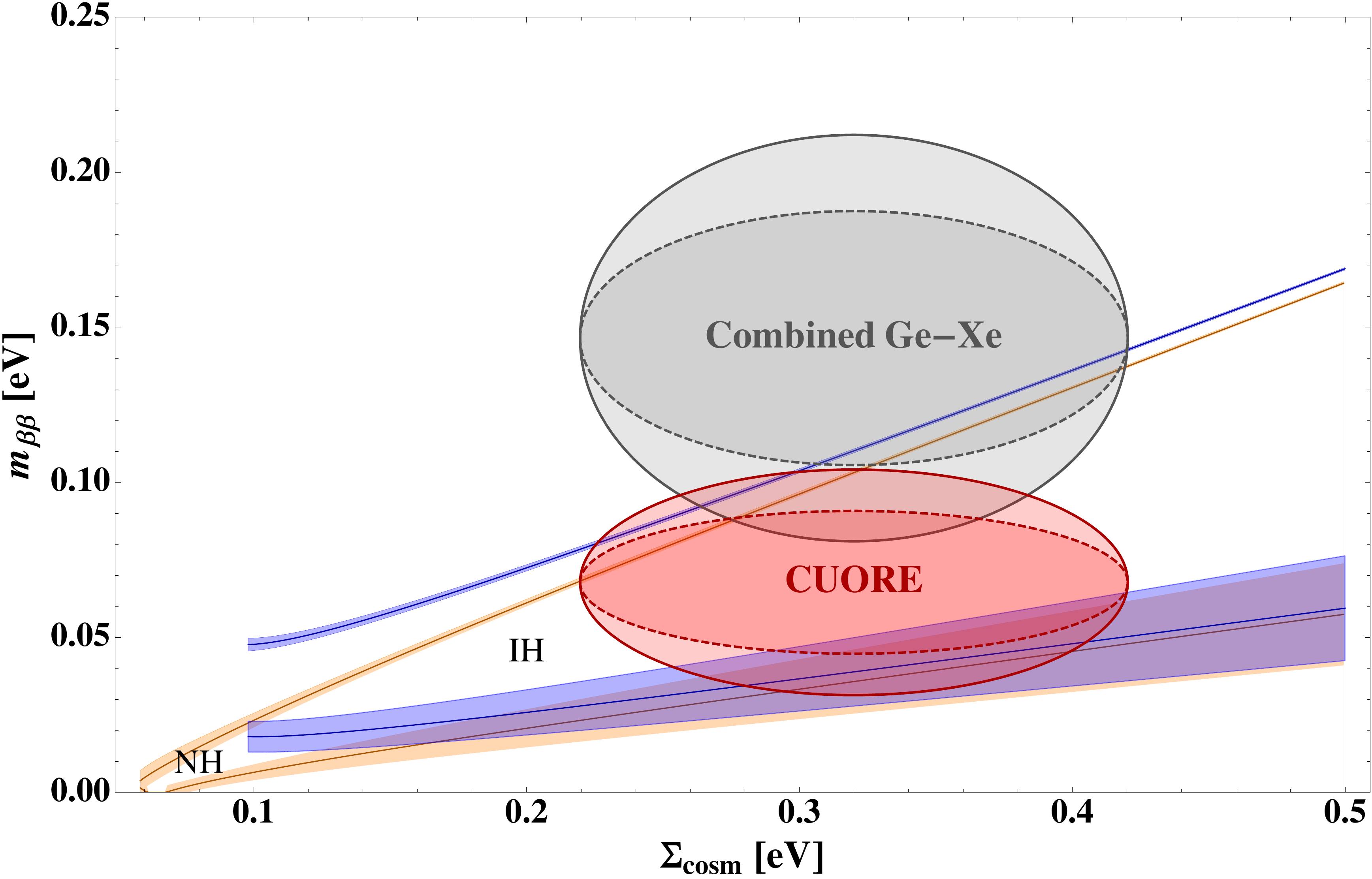}
		\caption{Allowed regions for $\mbb$ as a function of the neutrino cosmological mass $\Sigma$. 
			The colored bands correspond to the $3\sigma$ regions for the  extremal values of $\mbb$ as a 
			function of the neutrino cosmological mass $\Sigma$.
			On the left, the two ellipses represent the 90\%\,C.\,L.\ allowed regions for the couple 
			$(\Sigma$\,;\,$\mbb)$ according to the experimental limits quoted in the text (Sec.\ \ref{sec:upper_bounds}).
			On the right, the two big (small) ellipses show the 90\%\,C.\,L.\ regions in which a positive 
			observation of \bb~could be contained, according to the experimental uncertainties and 5 (20) 
			actually observed events. In particular, they refer to two different cases: the observation of \bb~with a $\mbb$ 
			corresponding either to the Ge+Xe limit or to the CUORE expected sensitivity [Eq.\ \ref{eq:nuova}]. 
			See Sec.\ \ref{sec:measurement} for a more detailed discussion.}
		\label{fig:ell3}
	\end{figure*}

\subsection{Combination of cosmology and \bb~results}

	We study two different situations. In the former case (Sec.\ \ref{sec:upper_bounds}), we assume that no effect of mass 
	is observed, and we have upper bounds both on $\Sigma$ and $\mbb$. 
	We use the conservative limit on $\Sigma$ reported from Planck {in Eq.\ (\ref{eq:limits_sigma})}. 
	As regards $\mbb$, we take the current best limit coming from the Ge+Xe combination (Table~\ref{tab:mbb_bounds}), 
	and the one corresponding 
	to the expected CUORE sensitivity (here, CUORE is chosen just as an example of a next generation experiment).
	In the latter case (Sec.\ \ref{sec:measurement}), we assume that the claim for neutrino mass from cosmology is correct  
	and that \bb~is measured with a half-life corresponding to the lowest bound of $\mbb$ coming from the Ge+Xe combination 
	or from the expected CUORE sensitivity. The values of $\mbb$ for these two cases are again those in 
	Table \ref{tab:mbb_bounds}.
	
\subsubsection{First scenario: Upper bounds}
	\label{sec:upper_bounds}

	Let us suppose Gaussian distributions centered in zero both for $\Sigma$ and $\mbb$, with a standard deviation coming 
	directly from the experimental upper limit; namely,  we put $\Sigma^{\meas}=\mbb^{\meas}=0$ in
	Eq.\ (\ref{eq:likelihood}). By requiring a 90\%\,C.\,L., we obtain the elliptic allowed regions in the left panel 
	of Fig.\ \ref{fig:ell3}. This picture shows that even in the CUORE case, there is no chance of ruling out the \IH, 
	unless there will be a great improvement on the knowledge of $\Sigma$. However, the combination of the two 
	parameters allows us to improve significantly the exclusion region.

\subsubsection{Second scenario: Measurements}
	\label{sec:measurement}

	Now we assume that both $\Sigma$ and $\mbb$ are measured with nonzero values.
	While the error on the former parameter comes directly 
	from Eq.\ (\ref{eq:limits_sigma}), the one on the latter one requires further discussion.
	
	The error on $\mbb$ has at least two different contributions: one 
	is statistical and comes from the Poisson fluctuations on the observed number of events, while the other one comes 
	from the uncertainties on the nuclear matrix elements and the phase space factors. 
	We will refer to this last as the \emph{theoretical} contribution to the total uncertainty.
	If we assume to know exactly the detector features (i.\,e.\ the number of decaying nuclei, the efficiency and the 
	time of measurement), the uncertainty on $\taubb$ is only due to the statistical fluctuations of the counts:
	\begin{equation}
		\frac{\delta \taubb}{\taubb}=\frac{\delta N_{\mbox{\tiny S}}}{N_{\mbox{\tiny S}}}.
	\end{equation}
	 The statistical contribution to the determination of the parameters is in general large and cannot be neglected. 
	 By emphasizing this simple but important point in the discussion, we consider a  case closer to the actual 
	 experimental situation, improving on the more idealized case that has been treated by previous 
	 investigators, \cite{Simkovic&al:2013_CP}. 
	
	The statistical contribution to the total error is dominant up to about 20 signal events. 
	The theoretical error becomes the main contribution only if many events (more than a few tens) are detected.
	Note that a much greater would come by taking into account the error on $g_A$. We assume here that this 
	problem will be solved in some manner in the future, and concentrate on the discussion of the role of the
	statistical error.
	
	Using the procedure described in Eq.\ (\ref{eq:err_prop}) for the Ge+Xe case,
	we find an uncertainty on $\mbb$ of about $31\,\mbox{meV}$ for 
	5 observed events, which reduces to $24\,\mbox{meV}$ for 10 events. If we neglect the statistical uncertainty, e.\,g.\ 
	we put $N_{\mbox{\tiny events}}>150$, the uncertainty becomes $14\,\mbox{meV}$. This means that the Poisson 
	fluctuations effect 
	is absolutely not negligible. Similarly, repeating the same calculation for CUORE, we obtain an uncertainty of 
	$17\,\mbox{meV}$ 
	for 5 events, $13\,\mbox{meV}$ for 10 events, and $8\,\mbox{meV}$ for $N_{\mbox{\tiny events}}>150$. 
	Therefore, referring to Table \ref{tab:mbb_bounds}, we consider the following cases:
	\begin{equation}
		\begin{split}
			&\mbb = \left(0.15 \pm 0.01_{\mbox{\tiny theo}} \pm 0.03 \, (1)_{\mbox{\tiny stat}} \right) \,\eV %
			\mbox{ \small (Ge-Xe)} \\[+5pt]
			&\mbb = \left(0.07 \pm 0.01_{\mbox{\tiny theo}} \pm 0.02 \, (1)_{\mbox{\tiny stat}} \right) \,\eV %
			\mbox{ \small (CUORE)}
			\label{eq:nuova}
		\end{split}
	\end{equation}
	where the statistical uncertainty is considered in the case of 5 (20) observed events and we computed the total error 
	by adding 
	the two contributions in quadrature.
	As for $\Sigma$, we assume the \emph{aggressive} value of Eq.\ (\ref{eq:limits_sigma}).
	The results are shown in the right panel of Fig. \ref{fig:ell3}. 
	The implication of these errors is further discussed in the next section.

\section{Is it possible to probe Majorana phases?}
	\label{uevos}
	
	Now we assume the optimistic scenario of Sec.\ \ref{sec:measurement} and consider the question of whether it is 
	possible to measure Majorana phases.  
	More precisely, we discuss the possibility of distinguishing the maximum and the minimum values of 
	$\mbb$, Eqs.\ (\ref{talp1}) and (\ref{talp2}). 
	In the case of quasidegenerate neutrinos that can be explored by present experiments, this possibility is 
	closely connected with the chance to measure one Majorana phase.
		
	Let us consider the allowed regions of parameters of the right panel of Fig.\ \ref{fig:ell3}.
	This picture shows that if CUORE observes five events (larger ellipse, continuous line), 
	we will not be able to reach any firm conclusion 
	either on the mass hierarchy or on the Majorana phases.
	Interestingly, if \bb~were instead discovered with a $\mbb$ a little bit below the current best limit on Ge+Xe, 
	this could allow us to make some inference on the Majorana phases. But it is important to repeat that, 
	in order to state anything precise about $\mbb$ and the Majorana phases, the present uncertainty on the quenching of 
	the axial vector coupling constant has to be dramatically decreased.
	
	When we repeat the same exercise assuming an observed number of 20 events, we obtain the smaller ellipses in the 
	right plot of Fig.\ \ref{fig:ell3} (dashed lines). 
	In this case, a hypothetical observation coming from the combined limit of Ge+Xe would lead to an even more 
	precise inference on the Majorana phases whereas, in the CUORE case, we would be closer to knowing 
	something useful on the Majorana phases, 
	even if nothing could be said about the hierarchy.

\section{Summary and discussion}
   
	We explored the hypotheses that the ordinary neutrinos are Majorana particles and that their
	exchange dominates the \bb~transition rate. In particular, we updated the predictions from neutrino oscillations and 
	we discussed the primary role played by considerations of nuclear physics and, more specifically, by the 
	axial vector coupling constant of the charged-current interactions of the nucleons.
	
	We stressed the importance of better understanding the quenching of $g_A$ in nuclear medium. 
	If this turns out to be negligible, it will be possible to probe the 
	\IH~region with the next generation experiments. Conversely, if this coupling is maximally quenched,  
	it will be unlikely to be able to reach the minimum sensitivity required to probe the 
	\IH~region within the next 20 years. Even in the optimistic scenario that the 
	\bb~will be discovered, it will be difficult to extract information on the process from the measurement, 
	if this uncertainty persists.
		
	We  argued that a measurement or a bound from cosmology could have an important impact on the expectations on $\mbb$.
	Indeed, cosmology could be precious to understand (and possibly quantify) the actual quenching of $g_A$.
	For example, if  the claim from cosmology of~\cite{Battye&Moss:2014} 
	were correct and if the future experiments 
	measured the \bb, we would conclude that the quenching effect is small or absent.
	
	We critically discussed the chances of measuring the
	Majorana phases, by quantifying the obstacles and by
	assessing the role of realistic experimental uncertainties.
	We showed that, at present, such a 
	measurement is really challenging, even in the most optimistic assumption on the quenching of the axial vector 
	coupling constant. 
	
	From the above discussion, further theoretical improvements and dedicated plans of measurements seem to be 
	necessary to clarify the expectations and decrease the uncertainties from nuclear physics. 
	 

	\begin{acknowledgments}
		We gratefully acknowledge extensive discussions with Prof.\ F.\ Iachello that stimulated this study.
		We thank the CUORE group at LNGS for precious discussions and important advice, and 
		Prof.\ S.\ Matarrese for explanations concerning the role of neutrinos in cosmology.
		We thank F.\ Nesti for helping us with the software part. 
		Finally, S.\ D.\ and S.\ M.\ are grateful to their colleagues at the GSSI for stimulating conversations and 
		friendly support. 
	
		Preliminary results were presented at La Thuile 2014, February 2014, and at IFAE 2014, LNGS, April 2014.
	\end{acknowledgments}

\appendix*
\section{\uppercase{Statistical procedures}}
	\label{app:stat}

\subsection{Combination of measurements}
	
	Let us consider the case of different neutrinoless double beta decay experiments using the same nucleus and quoting 
	the bounds on the half-life
	$\taubb > \taubbiso{i}\,\exper$ at the same confidence level, where $i=1,2,3,\dots$\,.
	A simple way to combine them is to suppose that the corresponding rates
	$\Gamma < \Gamma_i \equiv \ln 2 \cdot \hbar/\taubbiso{$i$}$ are Gaussian distributed in all the detectors, 
	namely the probability of observing 
	a rate within the interval $[\Gamma,\Gamma+d\Gamma]$ is $dP_i \propto  \exp [- \Gamma^2/2 \, \Gamma_i^2] \, d\Gamma$.
	This is the same as saying that the number of signal events is zero up to Gaussian fluctuations. 
	In this case, the combined Gaussian bound is 
	$\Gamma^{-2}_{\mbox{\tiny gaus.}}=\sum_i \Gamma^{-2}_i$. Therefore, we get the combined bound for the half-life 
	simply as 
	\begin{equation}
		\taubb = \sqrt{\sum_i (\taubbiso{$i$}\,\exper)^2}.
	\label{eq:comb_t}
	\end{equation}
	The described procedure has the advantage of being simple and generally conservative, although we remind the reader that 
	it should be validated in actual situations.
	
	The combination of results from different nuclei is more delicate and
	depends on the uncertain matrix elements. An elaborate procedure is
	discussed in \cite{Bergstrom:2012}. Our main goal is to outline the biggest factor
	of uncertainty, namely the dependence of the results upon $g_A$. Thus,
	working in the same hypotheses mentioned above, and assuming that the
	(relative) matrix elements are known precisely,
	we immediately obtain the bound on the mass relevant to the double beta decay:
	\begin{equation}
		\frac{1}{\mbb} = \left[
			\sum_{i=\mbox{\tiny \ce{Ge},\ce{Te}, \dots}}  
			\left( \frac{\taubbiso{$i$}\,\exper\, \Gi\, \mathscr{M}_i^2}{m_\el^2} \right)^{2}
		\right]^{1/4}.
		\label{eq:comb_xege}
	\end{equation}
	This is consistent with the theoretical expression of the half-life in the
   $i$th experiment, as given in Eq.\ (\ref{eq:tau}), and coincides with 
   Eq.\ (\ref{eq:comb_t}) for the same nuclear species.

\subsection{Error propagation}

	For any choice of the Majorana phases, the massive parameter that 
	regulates the \bb~can be thought as $M(m,\bm{x})$. It is a function of a mass 
	$m$ and of certain other parameters $\bm{x}$ that are determined by oscillation experiments up to their experimental 
	errors: $x_i\pm \Delta x_i$.
	
	Whenever we used maximal or systematic uncertainties from the literature, 
	we decided to interpret them as the semiwidths of flat 	
	distributions in order to propagate their effects in our calculations.
	Then, we considered the corresponding standard deviations 
	as Gaussian fluctuations of the parameters around the given values.
	
	For any fixed value of $m$ and for the other parameters set to their best fit values $x_i$, 
	we can attach the following error to $M$:
	\begin{equation}
		\left. \Delta M \right|_m =\sqrt{\sum_i \left( \frac{\partial M}{\partial x_i} \right)^2 \Delta x_i^2}.
	\label{eq:err_prop}
	\end{equation}
	If we want to consider the prediction and the error for a fixed value of another massive parameter
	$\Sigma(m,\bm{x})$, we have to vary also $m$, keeping
	$\delta \Sigma = \partial \Sigma/\partial m\, \delta m + \partial \Sigma/\partial x_i\, \delta x_i =0$. 
	Therefore, in this case we find:
	\begin{equation}
		\left. \Delta M \right|_\Sigma =
		\sqrt{\sum_i \left( \frac{\partial M}{\partial x_i} -
		\frac{{\partial \Sigma}/{\partial x_i}}{{\partial \Sigma}/{\partial m} } \,
		\frac{\partial M}{\partial m} \right)^2 \Delta x_i^2}.
	\end{equation}
	Of course, we calculate $m$ by inverting the equation $\Sigma(m,\bm{x})=\Sigma$. 
	(Here, the symbol $\Sigma$ denotes the function and also its value.
	However, this abuse of notation is harmless in practice.) 
	
\subsection{Confidence intervals}

	The likelihood  $\mathcal{L}$
	for the simultaneous observation of $\Sigma$ and $\mbb$,  Gaussian distributed variables with 
	uncertainties $\sigma(\Sigma^{\meas})$ and $\sigma(\mbb^{\meas})$, respectively is proportional to
	\begin{equation}
		\exp\!\left[
		 -\frac{(\Sigma-\Sigma^{\meas})^2}{2\sigma(\Sigma^{\meas})^2} 
		\right]
		\exp\!\! \left[ 
		-\frac{(\mbb - \mbb^{\meas})^2}{2\sigma(\mbb^{\meas})^2} \right].
	\label{eq:likelihood}
	\end{equation}
	This corresponds to the usual $\chi^2$: 
	\begin{equation}
		\chi^2=\frac{(\Sigma-\Sigma^{\meas})^2}{\sigma(\Sigma^{\meas})^2}+\frac{(\mbb - \mbb^{\meas})^2}{\sigma(\mbb^{\meas})^2}.
	\end{equation}
	The definition of the confidence intervals has to take into account the presence of 2 degrees of freedom.  
	Indicating with C.\,L.\ the desired confidence level, we have:
	\begin{equation}
		\text{C.\,L.} = \iint_{\chi^2 < \chi_0^2}   
		\mathcal{L}(\Sigma,\mbb)\ d\Sigma\ \, d\mbb .
	\end{equation}
	Thus, rescaling the variables of integration and integrating the angular coordinate, we have
	\begin{equation}
		\chi^2_0 = -2\,\ln (1- \text{C.\,L.})
	\end{equation}	
	which defines the value for $\chi^2$ corresponding to the assigned confidence level C.\,L.\,.

\bibliography{ref}
	
\end{document}